# BAYESIAN PHOTONIC ACCELERATORS FOR ENERGY EFFICIENT AND NOISE ROBUST NEURAL PROCESSING

George Sarantoglou, Adonis Bogris, Charis Mesaritakis and Sergios Theodoridis

*Abstract*— Artificial neural networks are efficient computing platforms inspired by the brain. Such platforms can tackle a vast area of real-life tasks ranging from image processing to language translation. Silicon photonic integrated chips (PICs), by employing coherent interactions in Mach-Zehnder interferometers, are promising accelerators offering record low power consumption and ultra-fast matrix multiplication. Such photonic accelerators, however, suffer from phase uncertainty due to fabrication errors and crosstalk effects that inhibit the development of high-density implementations. In this work, we present a Bayesian learning framework for such photonic accelerators. In addition to the conventional log-likelihood optimization path, two novel training schemes are derived, namely a regularized version and a fully Bayesian learning scheme. They are applied on a photonic neural network with 512 phase shifters targeting the MNIST dataset. The new schemes, when combined with a pre-characterization stage that provides the passive offsets, are able to dramatically decrease the operational power of the PIC beyond 70%, with just a slight loss in classification accuracy. The full Bayesian scheme, apart from this energy reduction, returns information with respect to the sensitivity of the phase shifters. This information is used to de-activate 31% of the phase actuators and, thus, significantly simplify the driving system.

## INTRODUCTION

Artificial neural networks (ANN) include a wide bio-inspired class of algorithms that is capable of learning and analyzing vast amount of data without being grounded on explicit instructions [1]. ANNs have allowed the development of a large area of novel information processing schemes that range from machine vision to language translation and audio processing [1]–[4]. The first ANNs have been implemented in traditional Von Neumann devices. However, as the number of available data is increased along with the need for fast processing, such computational paradigms suffer from an inherent problem, which is the separation between the central processing unit (CPU) and the memory unit [5]. To this end, a great amount of research nowadays focuses on new hardware architectures that break free from these limitations. Graphical Processing Units (GPUs) [6], application-specific integrated circuits (ASICs) [7], neuromorphic chips such as IBM's True North [8] and Intel's Loihi [9], have demonstrated considerable improvements in energy efficiency and processing speed.

In this area of study, photonic platforms have emerged as attractive candidates for analog ANNs. Photonics, in contrast to electronics, possess attributes such as ultrafast processing, wave division multiplexing assisted parallelism and high wall-plug efficiency [10]. In particular, photonic integrated silicon platforms are highly desirable schemes due to their low physical footprint and easy co-integration with electronic schemes. A special silicon photonics implementation based on coherent interaction of light through Mach Zehnder interferometers (MZI) showed significant energy consumption improvement when matrix – vector multiplications are considered [11]. This hardware paradigm acts as a photonic accelerator that implements ultrafast and power efficient weight matrices for ANNs. It consumes energy at the scale of fJ per floating point operation (FLOP) compared to 100 pJ per FLOP in conventional GPUs. A mesh of MZIs is configured based on the training data by applying the correct phase shifter values, so as to coherently modulate the incoming signal. Thus, in contrast to electronics the information processing itself consumes no power, as it entails only the propagation of the incoming light through the silicon photonics structure.

Although this functionality has been demonstrated with a dedicated photonic hardware that implements a feedforward mesh, it can be also supported by generic programmable gate arrays that are analogous to FPGAs in electronics. Such devices, known as field programmable photonic gate arrays (FPPGA), can be used to implement, apart from photonic accelerators for linear operations, other photonic components such as filters based on micro-ring resonators and MZIs [12], [13]. These multipurpose devices can be used to tackle a wide range of different tasks ranging from dispersion compensation in optical fibers [14] to quantum linear operations [15].

*Manuscript received February 11, 2022. This work was supported by the EU H2020 NEoteRIC project under grant agreement 871330. *(Corresponding author: G. Sarantoglou).*

George Sarantoglou and Charis Mesaritakis are with the Department of Information & Communication Systems Engineering, University of the Aegean, 83200 Karlovassi Samos, Greece (e-mail: gsarantoglou@aegean.gr, e-mail: cmesar@aegean.gr ).

Adonis Bogris is with the Department of Informatics and Computer Engineering, University of West Attica, 12243 Athens, Greece (e-mail: abogris@uniwa.gr ).

Sergios Theodoridis is with the is with the Department of Informatics and Telecommunications, National and Kapodistrian University of Athens, Athens, 15784 Attica, Greece and with the Electronics Systems Department, Aalborg University, Denmark (e-mail: stheodor@di.uoa.gr ).

The major problem in both FPPGAs and dedicated forward meshes is the uncertainty in the phases of the MZIs [16], when large scale matrices are targeted through highly integrated devices. Such uncertainties originate mainly from fabrication errors, dynamic sources of error and the bit precision of the voltage sources [11], [16], [17]. In conventional training of photonic accelerators, a loss function is minimized that returns the optimum phase shifter values to be set by the actuators. In order to apply optimally and precisely these phase shifter values, various techniques have been proposed in the published literature [17], [18].

In this work, we present an offline training scheme that considers the uncertainty in tuning the phases. In particular, we present a Bayesian treatment of the photonic accelerators, where, instead of defining through training the optimum phase shifter values, we define for each phase shifter a parametric probability distribution function (PDF), which is optimized by updating at every iteration its variational parameters. Through the Bayesian procedure, beside the correct phase shifter values, quantified information concerning the robustness of each phase shifter to phase deviations is provided. This information can be utilized to develop novel algorithms for the control and tuning of photonic accelerators, that potentially lead to increased noise robustness as well as to increased power efficiency. This work is inspired by advancements in the field of Bayesian Neural Networks (BNN) [19]–[21], [23]. BNNs are grounded on Bayesian statistics that treat the unknown model parameters as random variables. Randomness in Bayesian learning encodes our uncertainty with respect to the values of the unknown system weights. Thus, the task of learning comprises the inference of the associated PDFs. Usually, these are PDFs that are parameterized in terms of a set of related parameters, which are the ones to be optimized during training. The learned PDFs are known as the posterior PDFs, to contrast the prior PDFs, which are their initial estimates. The prior PDF, associated with a weight, is equivalent to a regularization term that biases the solution in the parameter space, where optimized values are searched [33]. Such a representation scheme yields important benefits such as immunity to overfitting [33], [20], low memory footprint [19], [21], [22] and simple weight pruning algorithms [21], [22]. These benefits illustrate that BNNs are considerably well tailored for hardware ANNs, where uncertainty with respect to the weights is a prominent issue [23]. Regarding hardware based BNNs, a spintronics system [24] has been proposed as well as an FPGA [25] implementation aiming at accelerated and energy efficient training and inference. Since on-chip training in PICs is yet in its infancy [23], we present an offline BNN training algorithm for photonic neural networks, whose benefits are harnessed during the inference stage.

The paper is organized as follows: in Section 2, we give a short overview of a subclass of photonic accelerators that instead of an arbitrary weight matrix implements a unitary weight matrix. Afterwards, we provide a Bayesian description of our system. First, this allows us to extract a method known as $\mathcal{L}_2$ regularization and the total Bayesian training method for photonic accelerators. In Section 3 we present the simulated photonic neural network that targets the MNIST dataset. The neural network is trained offline and the resulting values are loaded on chip. The case of thermal actuators is considered. We consider the fabrication errors, the thermal crosstalk effect and the quantization during inference. We describe a novel phase tuning technique based on the information provided by the training algorithm that allows the de-activation of thermal actuators. In Section 4, we present the simulation results. First, the bare $\mathcal{L}_2$ method is presented and afterwards the full Bayesian method. In Section 5, we provide some final comments with respect to the general prospects of the Bayesian framework in optical neural networks.

## II. THE BAYESIAN FRAMEWORK FOR OPTICAL ANNs

### A. The Photonic Accelerator

A photonic accelerator is an optical hardware implementation that performs a matrix – vector multiplication. By defining a physical node that is composed by an MZI and an external phase shifter and arranging such nodes in specific geometries, a matrix $U_{N\times N} \in \mathcal{C}^{N\times N}$ can be implemented [26], [27]. An example of an 8x8 photonic accelerator, implemented with a rectangular geometry is shown in Fig.1. Since this system is reciprocal, the matrix that is implemented is a unitary matrix. Although more complex mesh structures can implement any arbitrary matrix through the singular vector decomposition [11], unitary matrices are particularly appealing for photonic neuromorphic computing since they achieve similar performance while demanding lower physical footprint [17], [28].

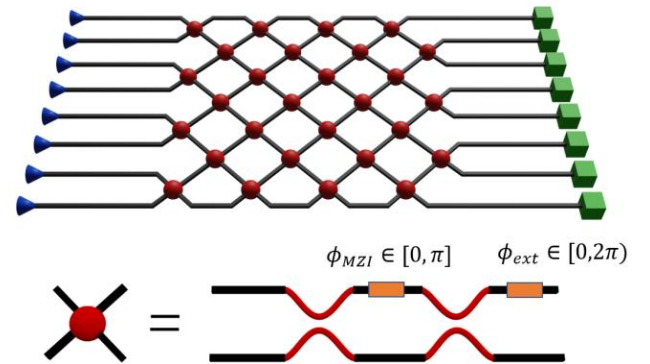

**Fig. 1**. The rectangular geometry for an 8x8 unitary matrix [27]. Each node represents a MZI followed by an external phase shift.

The basic block of a photonic unitary matrix is a physical node that consists of a MZI with an external phase shifter, whose transfer function is [28]:

$$U_{2\times 2}(\phi_{ext}, \phi_{MZI}) = ie^{i\phi_{MZI}/2} \begin{bmatrix} e^{i\phi_{ext}}\sin(\phi_{MZI}/2) & e^{i\phi_{ext}}\cos(\phi_{MZI}/2) \\ \cos(\phi_{MZI}/2) & -\sin(\phi_{MZI}/2) \end{bmatrix} \quad (1)$$

Here, $\phi_{ext} \in [0, 2\pi)$, $\phi_{MZI} \in [0,\pi]$ represent the phase of the external delay and the phase difference between the two arms of an MZI. In order to implement a $U_{N\times N}$ unitary matrix, $N(N-1)/2$ such nodes are required along with $N$ additional phase delays [26], [27]. Since a node is described by two phases, the whole unitary matrix is regulated by $N^2$ phase shifters. Supposing an ANN with $L$ neural layers, for the unitary weight matrix of the $l$-th layer $U^l_{N\times N}$ a vector $\phi_{1\times N^2}$ which contains all its phase shifter values must be calculated. The propagation of the electric field $X_l \in \mathcal{C}_{1\times N}$ through a unitary layer is given by

$$X_{l+1} = f_l(U^l X_l) = f_l(Z_l) \qquad (2)$$

The linear part $Z_l = U^l X_l$ is the matrix-vector multiplication that is implemented simply by propagating the light through the silicon mesh structure. The non-linear part $f_l(Z_l)$ can be implemented either by detecting the optical output $Z_l$ and processing it at the digital domain [11] or by more efficient electro-optic schemes such as the square law of the photodiode [29] or more specialized electro-optic schemes [30].

For notational simplicity, it is assumed that all the phase shifter values of the neural network are accumulated in a vector $\phi \in \mathcal{R}^{N_\phi}$, where $N_\phi$ is the total number of phase shifters that need to be tuned so as to configure all the unitary matrices $U^l_{N\times N}, l = 1,2 \dots, L$. The goal of the resulting optical neural network is to configure in an optimum way the vector $\phi$, so as to achieve the lowest classification error.

*B. The Bayesian framework*

Let $\mathcal{D} = \{(x_1,t_1),(x_2,t_2),\dots,(x_{N_d},t_{N_d})\}$ be the data set, that consists of $N_d$ input and target pairs. A parametric stochastic model $p(\mathcal{D}|\phi) = \prod_{i=1}^{N_d} p(t_i|x_i,\phi)$ maps the inputs $x$ to the targets $t$. This PDF is known as the likelihood and describes how reliably the dataset $\mathcal{D}$ is represented by a parametric model described by $\phi$. The Bayes rule states that

$$p(\phi|D) = \frac{p(D|\phi)p(\phi)}{p(D)} \qquad (3)$$

Here $p(\phi|D)$ is the posterior, since it describes the phase shifter values $\phi$ given the data $\mathcal{D}$. The term $p(\phi)$ is the prior, since it describes $\phi$ without any knowledge about $\mathcal{D}$. The term $p(D)$ is a PDF that describes the data set.

A conventional strategy is to find a vector $\phi_{MAP}$ that maximizes the posterior $p(\phi|D)$. This vector is known as the maximum a posteriori (MAP) estimator [31] and it expresses the most reliable choice of parameters given the information that is disclosed in $D$. Since the denominator in $p(\phi|D) = p(D|\phi)p(\phi)\mathbin{/}p D$ (3) does not depend on $\phi$, only the nominator $p(D|\phi)p(\phi)$ needs to be maximized.

The term $p(\phi)$ expresses the initial belief about the phase shifter vector $\phi$ before the observation of the data $D$. Under a Gaussian assumption, the vector $\phi$ is described by the following prior:

$$p(\phi) = \prod_{i=1}^{N_\phi} \frac{1}{\sqrt{2\pi}(\sigma_p)_i} \exp\left[-\frac{1}{2}\left(\frac{\phi_i - (\mu_p)_i}{(\sigma_p)_i}\right)^2\right] \qquad (4)$$

The components of the vector $\phi$ are regarded as independent random variables. Each component $\phi_i$ is described by a Gaussian with mean value $(\mu_p)_i$ and standard deviation $(\sigma_p)_i$, where $\mu_p$ and $\sigma_p$ are vectors. The values of these hyperparameters can be arbitrarily chosen by the user.

Maximizing $p(D|\phi)p(\phi)$ is equivalent to minimizing the relation $-\ln[p(D|\phi)p(\phi)]$ with respect to $\phi$. Consequently, $\phi_{MAP}$ is the value that minimizes:

$$\mathcal{L}_{MAP} = -\ln p(D|\phi) - \ln p(\phi) \qquad (5)$$

Neglecting the second term and focusing on the first term in (5), if the targets are described by a Gaussian distribution $t \sim p(D|\phi) = \mathcal{N}(X_{L+1}(\phi),x), \Sigma)$, where $X_{L+1}$ is the output of the last neural layer, then the minimization of the first term is equivalent to the minimization of the mean squared error loss function [31]. If the targets are described by a Categorical distribution $t \sim Cat(X_{L+1}(\phi,x))$, then the minimization of the same term is equivalent to the minimization of the categorical cross entropy cost function [31].

However, if the second term is also considered in (5), the minimization is equivalent to minimizing the term

$$\mathcal{L}_R(\phi) = \frac{1}{2}\sum_{i=1}^{N_\phi}\left(\phi_i - (\mu_p)_i\right)^2 / (\sigma_p)_i^2 \propto -\ln p(\phi) \qquad (6)$$

In machine learning literature, this term is known as $\mathcal{L}_2$ regularization [31]. By choosing a vector $\sigma_p$, the training algorithm punishes solutions of $\phi$ that are distant from the mean value $\mu_p$. For example, the probability of observing variables that deviate from the mean by more than $2\sigma_p, 3\sigma_p, 4\sigma_p$ and $5\sigma$ are $0.046, 0.003, 6\times 10^{-5}$ and $6\times 10^{-7}$, respectively.

The regularization term can be exploited to tackle the thermal crosstalk effect. In particular, initially, no voltage is applied on the phase shifters of the PIC, hence ideally $\phi_{MZI}$ should be zero, whilst $\phi_{ext}$ is determined by the length of the waveguide in Fig. 1 and the refractive index of the propagating mode. However, due to fabrication errors, the initial value of both phases is random and depends on the exact structural imperfections of the fabricated PIC. Thus, when no voltage is applied at the actuators, the phase shifters have random values $\phi_{offset} \in \mathcal{R}^{N_\phi}$. Power consumption and thermal crosstalk both increase proportionally with the value of voltage required per MZI so as to tune each value from the initial phase offsets $\phi_{offset}$ to the phase vector $\phi_{MAP}$ dictated by training. Thus, the regularization method is important with respect to the energy consumption and the thermal crosstalk on the chip. By restricting through $\sigma_p$ the $\phi_{MAP}$ vector close to the vector $\mu_p = \phi_{offset}$, the energy needed to tune the phase shifters is decreased. The classification error and the regularization error defined in (6), are competing when $\sigma_p$ is low since this vector restricts the search space for the algorithm.

In literature [11], [17], most authors do not consider the regularization term and focus solely on the minimization of the first term in (5). This is equivalent to setting all the elements of $\sigma_p$ equal to infinity. Then, $\mathcal{L}_R$ becomes zero and (5) becomes:

$$\mathcal{L}_{MAP} = -\ln p(D|\phi) \quad (7)$$

By using (7) instead of (5), a higher accuracy can be achieved on the training data, since the searching space of the optimum $\phi_{MAP}$ is not restricted by the prior (6). However, the goal in a machine learning is the generalization performance, that is how the designed network behaves on test data that did not participate in the training. If no regularization is used, the designed network is prone to overfitting [31]. Additionally the thermal crosstalk effect is more prominent, since this scheme does not consider the minimization of heat generated by the thermal actuators and therefore the performance is degraded by thermal gradients proportional to the magnitude of the applied phase shifts [32].

The MAP method returns a specific $\phi_{MAP}$ vector. However, the most useful solution would be the posterior $p(\phi|D)$ itself, since this provides the full statistical information about the values to be learned. In this case, when the phases of the chip are tuned, one is not restricted to a particular vector $\phi_{MAP}$. A vector $\phi$ can be chosen that belongs in an $N_\phi$-th dimensional region with high $p(\phi|D)$. Such a vector is a reliable choice for the parametric model that describes the data $\mathcal{D}$.

To this account, the variational inference method is used [19]. In particular, a PDF $q_\theta(\phi)$ is defined, where $\theta$ is a set of variational parameters, that approximates as well as possible the true posterior $p(\phi|D)$. This can be achieved by computing the optimum $\theta^*$ parameter values that minimize the Kullback – Leibler divergence between $q_\theta(\phi)$ and $p(\phi|D)$ [20]. This is equivalent to minimizing the following loss function [20]:

$$\mathcal{L}_B(\phi, \theta) = \mathbb{E}_{q_\theta(\phi)}[\mathcal{L}_E(\phi)] + \mathbb{E}_{q_\theta(\phi)}[\mathcal{L}_R(\phi)] - \mathcal{H}(q_\theta(\phi)) \quad (9)$$

The first two terms describe the mean value of the conventional loss function and the regularization function with respect to the $q_\theta(\phi)$ distribution, respectively. The third term is the entropy of $q_\theta(\phi)$. For the shake of simplicity, instead of sampling multiple instances of $\phi$ from the posterior $q_\theta(\phi)$ to evaluate the mean values in (9), only a single sample from the posterior is used. We regard that $q_\theta(\phi)$ is a product of Gaussian distributions:

$$q_\theta(\phi) = \prod_{i=1}^{N_\phi} \frac{1}{\sqrt{2\pi\sigma_i^2}} \exp\left[-\frac{1}{2}\left(\frac{\phi_i - \mu_i}{\sigma_i}\right)^2\right] \quad (10)$$

Here $\theta = (\mu, \sigma) \in \mathcal{R}^{2N_\phi}$ with $\mu, \sigma \in \mathcal{R}^{N_\phi}$. Thus, a sample from the posterior is $\phi = \mu + \epsilon \circ \sigma$, where $\epsilon \sim \mathcal{N}(0, I)$. The training procedure aims to find the optimum vector $\theta^*$ which minimizes the loss function $\mathcal{L}_B$. Each step of the optimization procedure is as follows [20]:

1) Sample $\epsilon \sim \mathcal{N}(0, I)$.
2) Let $\sigma = \ln(1 + e^{\rho_\sigma})$.
3) Draw $\phi = \mu + \sigma \circ \epsilon$.
4) Compute $\mathcal{L}_B(\phi, \mu, \sigma) = \mathcal{L}_C(\phi) + \mathcal{L}_R(\phi) + \ln q_\theta(\phi)$
5) Find the gradients with respect to $\mu$
$$\frac{d\mathcal{L}_B}{d\mu} = \frac{\partial \mathcal{L}_B}{\partial \mu} + \frac{\partial \mathcal{L}_B}{\partial \phi}$$
6) Find the gradients with respect to $\rho_\sigma$
$$\frac{d\mathcal{L}_B}{d\rho_\sigma} = \frac{\partial \mathcal{L}_B}{\partial \rho_\sigma} + \frac{\epsilon}{1 + \exp(-\rho_\sigma)} \frac{\partial \mathcal{L}_B}{\partial \phi}$$
7) Update with gradient descent the parameters $\mu, \rho_\sigma$
$$\mu \leftarrow \mu - \eta d\mathcal{L}_B/d\mu$$
$$\rho_\sigma \leftarrow \rho_\sigma - \eta d\mathcal{L}_B/d\rho_\sigma$$

In step 2, the standard deviations $\sigma$ are expressed in terms of the variational parameters $\rho_\sigma \in \mathcal{R}^{N_\phi}$, through the softplus function $f(x) = \log(1 + exp(x)) > 0, \forall x \in \mathcal{R}$. Thus, by updating $\rho_\sigma$ it is reassured that the standard deviation always holds positive values $\sigma > 0$. At this point it should be stressed that the gradients $\partial \mathcal{L}_B / \partial \phi$ are the exact same gradients derived by the back-propagation algorithm [20]. The gradients $\partial \mathcal{L}_B / \partial \mu, \partial \mathcal{L}_B / \partial \rho_\sigma$ can be computed analytically. The only trade-off of this method is that it has to compute the double parameters compared to the MAP estimation, however the gain is that through $\theta^*$ the user has valuable information about the sensitivity of all phases.

After the training procedure, each phase shifter is associated with a PDF as it is illustrated in Fig. 2. The returned vector $\theta^*$ provides information regarding the optimum mean value vector $\mu^*$ and standard deviation vector $\sigma^*$. Given that $q_{\theta^*}(\phi)$ is the approximation of the true posterior $p(\phi|D)$, the samples $\phi$ from the posterior that belong to the region $\mu^* \pm \sigma^*$ with the higher values of $q_{\theta^*}(\phi)$ are desired. In particular, the approximate posterior is used as it were the true posterior. Vectors $\phi$ that provide high $q_{\theta^*}(\phi)$ are the most reliable $\phi$ options given the observed data $D$ according to the approximate posterior. On this account, phase shifters with high standard deviation need much lower precision during the tuning procedure compared to phase shifters with lower standard deviation.

The consequence of these observations is that the network can be efficiently tuned, by taking advantage of the additional information about the significance of each phase shifter in order to achieve both high accuracy and the minimum consumption. In the presence of high phase uncertainties it is challenging to tune large-scale photonic meshes[23], and consequently such information is critical.

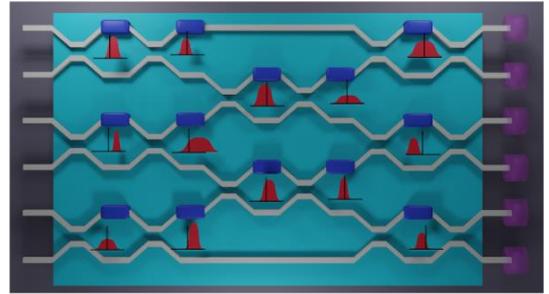

**Fig. 2.** The Bayesian description of a photonic accelerator. A Gaussian PDF is associated with each phase shifter.

### III. SIMULATION OF THE BAYESIAN PHOTONIC ACCELERATOR

The presented analysis has two goals. First, to highlight the importance of the $\mathcal{L}_2$ regularization method for the decrease of the power consumption and consequently the thermal crosstalk effect. The second goal is to show the potential of the full Bayesian method described by (9). Apart from power reduction this method provides additional information that can be used to de-activate thermal actuators.

In all cases, the system will be trained offline and the resulting phase vector $\phi$ will be loaded by applying on chip the phase deviation $d\phi = \phi - \phi_{measured}$. It is assumed that before training, a pre-characterization process takes place that provides the relation between the applied current and the induced phase shift for each actuator [13]. This method provides information regarding the phase offsets due to fabrication errors. Although such a process is rather complex as the scale of photonic accelerator increases, it needs to be performed only once for each device, whereas the benefits with regard to its real-time operation are considerable. Since MZIs at the scale of $300 - 600\ \mu m$ are simulated that are compatible with the first generation of photonic accelerators based on FPPGAs, it is regarded that the phase offsets are drawn by a uniform distribution between 0 and $2\pi$ [24]. Thus, the measured phase offsets are $\phi_{measured} = \phi_{offset}$.

Additionally, the quantization error is simulated for 8-bit precision thermal actuators. The resulting quantized vector is represented as $d\phi_q = [\phi - \phi_{measured}]_q$. Last, the thermal crosstalk effect is simulated by means of a thermal matrix $T \in \mathcal{R}^{N_\phi \times N_\phi}$ that is given by the relation (11) [24].

$$T = \begin{bmatrix} 1 & CT_{12} & \ldots & \ldots & \ldots & CT_{1N_\phi} \\ CT_{21} & 1 & & & & CT_{2N_\phi} \\ CT_{31} & & 1 & & & \vdots \\ \vdots & & & \ddots & & \vdots \\ \vdots & & & & \ddots & CT_{N_\phi-1,N_\phi} \\ CT_{N_\phi 1} & \ldots & \ldots & \ldots & \ldots & 1 \end{bmatrix} \quad (11)$$

Each term $CT_{ij}$ describes the portion of the phase shift that is caused on the phase shifter $i$ due to a thermal actuator acting on phase shifter j. A pessimistic scenario is considered where the non-diagonal terms $CT_{ij}$ are sampled by a uniform distribution $U(0, CT)$ with $CT$ expressing the thermal crosstalk coefficient [24]. In this case, phase shifters that are considerably distant in the mesh may influence one another. In reality the thermal coefficient ranges between 1-3% and it is a function of the distance[33]. This value can be decreased further in future implementations with the development of better thermal trenches [34]. Consequently, the resulting phase shift on the chip is:

$$d\phi_{chip} = \phi_{offset} + T[d\phi]_q \quad (12)$$

During inference the relation (12) is used. Since this work focuses solely on the phase error, sources of error such as the propagation loss and the photodetection error at the photodiodes are ignored.

With respect to the offline training process, all the loss functions will be considered. When minimizing either (5) or (9), the mean prior is set equal to $\mu_p = \phi_{measured}$ and the standard deviation of the prior is considered common for all the phase shifters and consequently it can be substituted in (4) by a scalar value $\sigma_p^s$, where the superscript stands for scalar. By equating the mean of the prior with the measured phase offsets, solutions that deviate significantly from the passive offsets are punished. Training with (7) is referred as the log-likelihood training scheme, with (5) as the regularization scheme and with (10) as the Bayesian scheme.

The log-likelihood and the regularization schemes return a vector $\phi_{MAP}$ and the applied phase shift is $d\phi = \phi_{MAP} - \phi_{measured}$. In the Bayesian scheme, since information about the robustness of the phase shifters is provided through the $\sigma$ vector, redundant actuators can be de-activated. In particular, if a phase offset $(\phi_{measured})_i$ is inside the region $\mu_i \pm \sigma_i$ then no thermal shift is needed and the thermal actuator is de-activated resulting in a phase shift $d\phi_i = 0$ rad. For other phase shifters the phase shift is $d\phi_i = \mu_i - u_i\sigma_i$. The terms $u_i \in [0, 1)$ can be tuned by the user. Higher $u_i$ translates to lower energy consumption since a lower phase shift $d\phi$ is applied. However, in this case the phase shifter might be more sensitive to dynamic phase errors. Given a choice $u_i$, a deviation equal to $(d\phi_i)_{noise} > (1 - u_i)\sigma_i$ is able to lead the phase outside the optimum region $\mu_i^* \pm \sigma_i^*$. The term $(d\phi_i)_{noise}$ originates from the thermal crosstalk effect, the precision error of the actuators, etc. The user can regulate $u_i$ according to whether high precision or lower power consumption is desired. Since thermal actuators are treated in this work, all terms $u_i, i = 1, \ldots, N_\phi$ are set equal to 0.9.

IV. SIMULATION RESULTS

The MNIST dataset is used to benchmark the three training schemes [37]. The simulated setup is presented in Fig. 3. Each feature which consists of 28 x 28 pixels is compressed at an 1x16 complex vector after the application of the 2D Fourier transform. The information lies at the center of the 2D spectrum and thus by keeping the 16 central coefficients, the essential information is maintained [30]. The optical ANN consists of 2 layers and its input consists of 16 complex fields that are equal to a normalized version of the 16 Fourier complex coefficients. The linear operation of each layer is associated with a trainable 16x16 unitary matrix. Thus, in total $N_\phi = 2 \times 16^2 = 512$ phases have to be tuned. The first layer (hidden layer) has the non-linear function $f(z) = |z|$, whereas the second layer (output layer) has the non-linear function $f(z) = |z|^2$. These functions can be implemented by using a simple incoherent detection scheme at the output of each layer. The dataset is divided in 60000 digits for training and 10000 digits for testing. Mini-batches of 50 features are used and the total training procedure spans for 25 epochs.

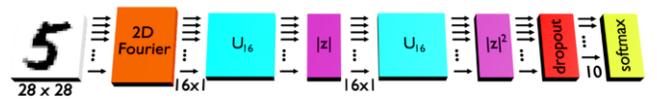

**Fig. 3.** The simulated setup for the MNIST dataset. A pre-processing change extracts from each 28x28 picture 16 complex coefficients that are inserted in a unitary neural network with 2 unitary matrices and two non-linear activation functions. At the end, a dropout layer extracts the 10 first

outputs and guides them to a softmax layer that predicts the correct class.

*A. The Regularization scheme*

First, the regularization scheme described by (5) is compared to the log-likelihood scheme described by (7). The value $\sigma_p^s$ of the prior is treated as a hyperparameter. For each examined value, 50 different phase offset vectors $\phi_{offset}$ are used corresponding to 50 different devices. For each device, the pre-characterization stage takes place, followed by the offline training procedure. The computed $\phi_{MAP}$ vector is applied on-chip according to (12).

First, only the power efficiency and accuracy are examined and therefore, the thermal crosstalk coefficient is set to 0 %, whereas the actuators with 8-bit precision are considered. In order to quantify the power consumption at the phase shifters, the following metric is introduced:

$$L_{norm} = \frac{1}{N_t}\sum_{k=1}^{N_t}\sum_{i=1}^{N_\phi} d\phi_i^k \qquad (13)$$

Here, $N_t = 50$ is the number of different devices and $d\phi_i^k$ is the phase modification that is applied on the i-th phase shifter of the k-th device. It is assumed that each phase shift consumes power equal to $10/\pi$ mW/rad [11]. By multiplying the metric $L_{norm}$ with this number, the total power consumption is derived. The resulting accuracy and energy consumption are depicted in Fig. 4. It can be seen, that the log-likelihood scheme is superior compared to the regularization scheme with regard to the accuracy. This result is natural, since the log-likelihood scheme focuses only on the classification error, whereas the regularization scheme tries to minimize both the classification error and the $\mathcal{L}_2$ term in (5). However, the sacrifice in accuracy is not considerable when a suitably high $\sigma_p^s$ is used. Here for $\sigma_p^s = 0.1$ rad the mean accuracy is 0.81 compared to 0.84.

On the other hand, the power efficiency demonstrates the clear benefit that is harnessed by using the regularization scheme. As the $\sigma_p^s$ term is reduced, the power consumption is decreased. Even for $\sigma_p^s = 0.1$ rad the metric is equal to 250 rad compared to 964 rad for the log-likelihood scheme indicating a 74 % power reduction. By a further lower $\sigma_p^s = 0.05$ rad, the accuracy is equal to 0.79 and the power metric is 151.5 rad indicating approximately an 84.2% power saving.

This reduction is particularly important in the case of the thermal actuators since apart from the power saving, it is linked to alleviation of the thermal crosstalk effect. In order to analyze this effect, for the regularization scheme the standard deviation of the prior $\sigma_p^s = 0.05$ rad so as to maintain a balanced trade-off between accuracy and power consumption. For each thermal coefficient, 100 different randomly chosen thermal matrices (11) are used and the presented results are the average inference accuracies. Since the scenario that is described by the thermal crosstalk matrix is very pessimistic, low thermal crosstalk coefficients are analyzed ranging from 0 to 1%. The comparison between the regularization and the log-likelihood scheme is presented in Fig. 5. For 0% thermal coefficient the accuracy of the log-likelihood is only slightly better than that of the regularization scheme as in Fig. 4. However, as the thermal crosstalk coefficient is increased the accuracy of the log-likelihood scheme falls abruptly, whereas the accuracy of the regularization scheme falls smoothly. This shows, that training with the $\mathcal{L}_2$ regularization term provides in this case more robust solutions compared to the log-likelihood scheme.

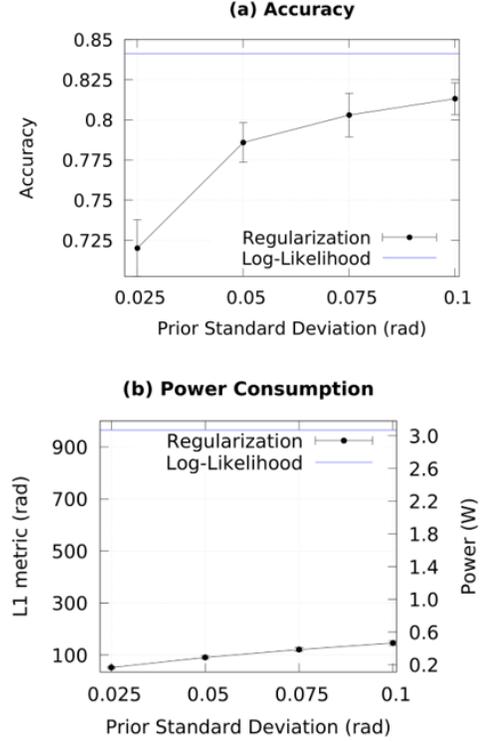

**Fig. 4.** The mean accuracy (a) and the mean power consumption (b) in the case of the MNIST dataset, for the regularization training scheme as a function of the standard deviation of the prior. This scheme is compared in these graphs with the log-likelihood training scheme.

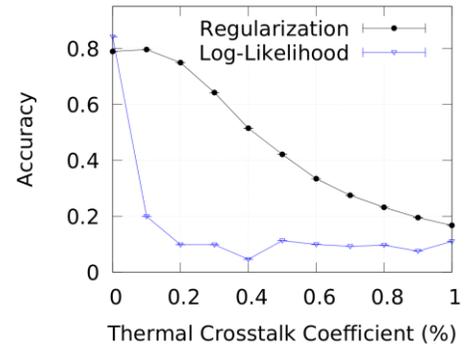

**Fig. 5** The accuracy for the solutions given by the regularization scheme and the log-likelihood scheme for the MNIST dataset as a function of the thermal crosstalk coefficient.

*B. The Bayesian scheme*

Next, the regularization scheme is compared to the Bayesian scheme. It must be stressed that the regularization term is already incorporated in (9) and consequently the Bayesian scheme can attain a similarly significant power reduction.

At this stage, the reduction in complexity of the driving system, by setting the voltage of thermal actuators to zero, is examined. Although, this effect can be achieved also by the regularization scheme for phase shifts below the limit defined by the quantization error, the information provided by the Bayesian method can be used to further decrease the complexity. The system will be trained offline with the Bayesian method presented in Section II.B and the actuators that are pointed to phase shifters with phase offsets inside the region $\mu \pm \sigma$ will be set to zero. The standard deviation of the prior is set at $\sigma_p^s = 0.1$ rad. As before, the choice $\mu_p = \phi_{offset}$ is made for the prior due to the pre-characterization stage.

In the histogram in Fig. 6, the standard deviations of the 512 phase shifters are presented. In this graph each bin has a 0.01 rad width. It can be seen from the subfigure that 200 phase shifters values are very sensitive to phase shifter perturbation with standard deviation $\sigma^* = 0.01 \pm 0.005$ rad. There are also 187 phase shifters in the range $\sigma^* = 0.02 \pm 0.005$ rad. Other phase shifters are more robust to deviations with higher standard deviation values. Moreover, there are also many phase shifters with standard deviations sparsely distributed up to 2 rad. Consequently, Fig. 6 provides a full picture about the optical ANN weights and their robustness to variations.

By exploiting this information, the offline training algorithm is repeated for different values of $\sigma_p^s$ ranging from 0.025 rad to 0.1 rad. In the number of thermal actuators that are set to zero versus the $\sigma_p^s$ parameter is presented. The quantization error is again equal to 8-bit and the thermal crosstalk coefficient is set to 0%. There is a 5% increase in the zero voltage actuators for the Bayesian case compared to the regularization procedure, thus further simplifying the driving system. The price to pay here is observed in the accuracy which slightly reduces in the case of the Bayesian procedure.

In order to clarify that the Bayesian framework enables the user to de-activate thermal actuators mainly through information provided by the training procedure, the standard deviation of the prior is fixed at $\sigma_p^s = 0.1$ rad and the training process is repeated for different bit precisions at the actuators ranging from 4 to 16 bits. The results are shown in Fig. 8.

Although, the regularization process and the Bayesian process share the same number of de-activated actuators at low bit precision, the reason behind this simplification is the minimum limit that is set by the quantization error. As that error is decreased by increasing the bit precision of the actuators, the number of de-activated actuators reduces for the regularization process, whereas in the case of the Bayesian scheme it reaches a plateau at 162 phase shifters (31% of actuators are de-activated). Therefore, as the quantization error is reduced, the Bayesian scheme is able to simplify the driving system based solely to its information about the sensitivity of phase shifters.

The classification accuracy for 4-bit precision was low for both schemes and equal to 0.22. For 8-bit and up to 16-bit, the accuracy was approximately equal to 0.81. In terms of the power consumption, both schemes return the same L1 norm (13). However, for high bit precision, in cases where the regularization scheme dictates small phase shifts, the Bayesian scheme by harnessing the additional information about the sensitivity of the phase shifter, allows us to de-activate the thermal actuators entirely, thus significantly simplifying the driving process.

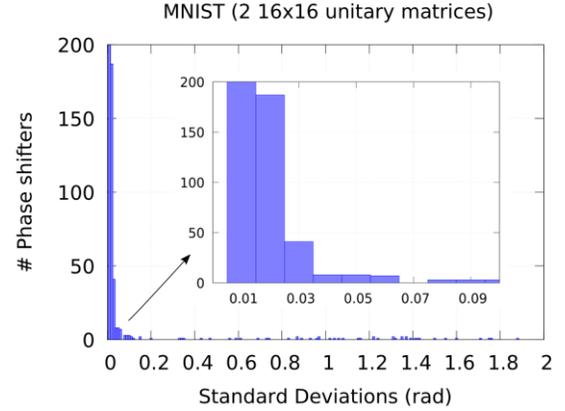

**Fig. 6.** The standard deviation of the 512 phase shifters as determined by the Bayesian training algorithm, needed to tackle the MNIST problem.

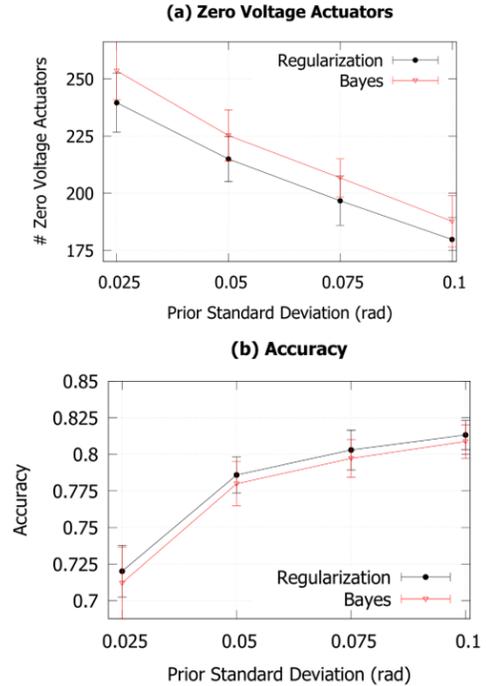

**Fig. 7.** The number of de-activated actuators (a) and the accuracy (b) versus the standard deviation of the prior for the regularization and the Bayesian process.

## V. DISCUSSION

The presented analysis focuses solely on thermal actuators that are used in the first generation of both dedicated forward meshes and FPPGAs implementing linear operations. The use of a regularization scheme when combined with a pre-characterization stage can dramatically decrease the power consumption compared to the conventional training schemes presented in literature. The Bayesian path, although it trains twice as many parameters, it provides important guidelines with respect to the sensitivity of the phase shifters. Since it incorporates the regularization process it is also beneficial to power consumption.

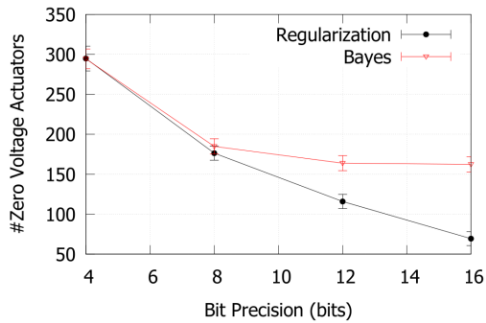

**Fig. 8.** The number of de-activated thermal actuators versus the bit precision.

In the case of next generation photonic accelerators which are based of volatile BTO phase shifters [35], phase change materials [36] or III-V materials [37], the power consumption is dramatically reduced and therefore the minimization of the power consumption is not as crucial as in the case of thermal actuators. Beside the fact that the Bayesian procedure will allow for power saving beyond 70 % even in these cases, the greatest gain would be the simplification of the electronics needed to drive the phase shifters, since redundant actuators are set to zero voltage. Such a benefit will be important as the size of photonic accelerators scales up.

Moreover, with respect to the Bayesian algorithm itself only the case of phase shifters described by Gaussian distributions is examined. It should be stated that the presented mathematical formalism allows the incorporation of additional auxiliary random variables, that can provide extra benefits for the simplification of the setup [21], [22]. It must also be considered that using more biologically plausible non-linear functions such as the winner takes it all layer, the robustness of a neural network can be enhanced by allowing for phase shifter values with as lower bit precision [22]. Lastly, although the Bayesian treatment of a silicon PIC based on Mach Zehnder interferometers has been presented, the aforementioned functionalities can be also incorporated in different hardware photonic platforms such as spatial [29] or time-delayed reservoir computing [38] implementations or full spiking neural networks [10].

## VI. CONCLUSION

We provide a full Bayesian treatment of silicon PIC coherent accelerators for neural processing in linear programmable photonic devices. Two methods are extracted from this formalism, the regularization method that focuses solely on power reduction and the full Bayesian method that in addition returns information about the sensitivity of all phase shifters. In terms of energy reduction, both a dramatic decrease in energy consumption beyond 70 % and robustness to thermal crosstalk noise are observed for a large-scale mesh with 512 phase shifters targeting the MNIST dataset. The information about the sensitivity of the phase shifters is also exploited during the inference process to simplify the driving electronics by disabling redundant thermal actuators. These results, render the Bayesian formalism significantly beneficial for energy efficient and computationally powerful large-scale optical neural networks implemented in any programmable silicon photonics device.

**George Sarantoglou,** received the Diploma degree in electrical and computer engineering from the University of Patras, Patras, Greece, on 2016. He is currently working towards his Ph.D. degree with the Department of Information and Communication Systems Engineering, University of the Aegean, Samos Greece. His Ph.D. thesis focuses on the experimental analysis and development of photonic processors for unconventional, bio-inspired information processing, targeting machine learning applications. His research interests include photonic systems for analog pattern recognition and multi-sensory applications.

**Adonis Bogris**, was born in Athens. He received the B.S. degree in informatics, the M.Sc. degree in telecommunications, and the Ph.D. degree from the National and Kapodistrian University of Athens, Athens, in 1997, 1999, and 2005, respectively. His doctoral thesis was on all-optical processing by means of fiber-based devices. He is currently a Professor at the Department of Informatics and Computer Engineering at the University of West Attica, Greece. He has authored or coauthored more than 150 articles published in international scientific journals and conference proceedings and he has participated in plethora of EU and national research projects. His current research interests include high-speed all-optical transmissions systems and networks, non-linear effects in optical fibers, all-optical signal processing, neuromorphic photonics, mid-infrared photonics and cryptography at the physical layer. Dr. Bogris serves as a reviewer for the journals of the IEEE and OSA.

**Charis Mesaritakis**, received the B.S. degree in Informatics, from the Department of Informatics & Telecommunications of the National & Kapodistrian University of Athens in 2004, the M.Sc. degree in microelectronics from the same department, whereas in 2011 and the Ph.D degree on the field of quantum dot devices and systems for next generation optical networks, in the photonics technology & optical communication laboratory of the same institution. In 2012 was awarded a European scholarship for post-doctoral studies (Marie Curie FP7-PEOPLE IEF) in the joint research facilities of Alcatel-Thales-Lucent in Paris-France, where he worked on intra-satellite communications. He has actively participated as research engineer/technical supervisor in more than ten EU-funded research programs (FP6-FP7-H2020) targeting excellence in the field of photonic neuromorphic computing, cyber-physical security, and photonic integration. He is currently an Associate Professor at the Department of Information & Communication Systems Engineering at the University of the Aegean, Greece. He is the author and coauthor of more than 80 papers in highly cited peer reviewed international journals and conferences, two international book chapters, whereas he serves as a regular reviewer for IEEE and OSA.

**Sergios Theodoridis**, (Life Fellow, IEEE) is currently a Professor Emeritus with the National and Kapodistrian University of Athens, Athens, Greece, and a Distinguished Professor with Aalborg University, Aalborg, Denmark. He is the author of the book Machine Learning: A Bayesian and Optimization Perspective (Academic Press, 2nd Edition, 2020), the coauthor of the best-selling book Pattern Recognition (Academic Press, 4th Edition, 2009), the coauthor of the book Introduction to Pattern Recognition: A MATLAB Approach (Academic Press, 2010), and the coeditor of the book Efficient Algorithms for Signal Processing and System Identification (Prentice Hall, 1993). His research interests span a wide range of areas in the intersection of signal processing and machine learning. He has received a number of best paper awards, including the 2014 IEEE Signal Processing Magazine Best Paper Award and the 2009 IEEE Computational Intelligence Society IEEE Transactions on Neural Networks Outstanding Paper Award. He was the recipient of the 2022 IEEE Signal Processing Society Norbert Wiener Award, the recipient of the 2017 EURASIP Athanasios Papoulis Award, the 2014 IEEE Signal Processing Society Carl Friedrich Gauss Education

Award, and the 2014 EURASIP Meritorious Service Award. He has served as the Vice President for the IEEE Signal Processing Society, as the President for the European Association for Signal Processing (EURASIP), and as the Editor-in-Chief for IEEE Transactions on Signal Processing. He is a Fellow of the IET and EURASIP and a Corresponding Fellow of the Royal Society of Edinburgh (RSE).